%% file: Margaroli_LHCP.tex
\documentclass[10pt]{article}
\usepackage{graphicx}

\def\Title#1{\begin{center} {\Large #1 } \end{center}}
\def\Author#1{\begin{center}{ \sc #1} \end{center}}
\def\Address#1{\begin{center}{ \it #1} \end{center}}

\newcommand\pubblock{\rightline{\begin{tabular}{l} Proceedings of the Second Annual LHCP\\ \pubnumber\\
         \pubdate  \end{tabular}}}

\newenvironment{Abstract}{\begin{quotation} \begin{center} 
             \large ABSTRACT \end{center}\bigskip 
      \begin{center}\begin{large}}{\end{large}\end{center} \end{quotation}}

\newenvironment{Presented}{\begin{quotation} \begin{center} 
             PRESENTED AT\end{center}\bigskip 
      \begin{center}\begin{large}}{\end{large}\end{center} \end{quotation}}

\def\Acknowledgements{\bigskip  \bigskip \begin{center} \begin{large}
             \bf ACKNOWLEDGEMENTS \end{large}\end{center}}

\input econfmacros.tex

\usepackage{color}

\newcommand\pubnumber{arxiv:1409.4208 }

\newcommand\pubdate{\today}

\def\affiliation{
On behalf of the CDF and D0 Experiments, \\
University of Rome Sapienza \& INFN Roma1 \\
P.zle Aldo Moro 5 00186 Rome, Italy}

\begin{document}

\large
\begin{titlepage}
\pubblock

\vfill
\Title{Higgs physics at the Tevatron  }
\vfill
\Author{Fabrizio Margaroli}
\Address{\affiliation}
\vfill
\begin{Abstract}
We show the latest results from the CDF and D0 collaborations on the study of the Higgs boson, stemming from the analysis
of the entire Tevatron Run\,II dataset. Combining the results of many individual analyses, most of which use the full data set available, an excess with a significance of approximately three standard deviations with respect to the Standard Model hypothesis is observed at a Higgs boson mass of 125\,GeV/$c^2$. The Tevatron unique environment allows in addition to study for the first time the spin-parity hypothesis of the Higgs boson in events where it decays to quarks. Within the current experimental uncertainties, the newly discovered boson behaves as expected by the SM in the fermionic sector.

\end{Abstract}
\vfill

\begin{Presented}
The Second Annual Conference\\
 on Large Hadron Collider Physics \\
Columbia University, New York, U.S.A \\ 
June 2-7, 2014
\end{Presented}
\vfill
\end{titlepage}
\def\thefootnote{\fnsymbol{footnote}}
\setcounter{footnote}{0}
%

\normalsize 


\section{Introduction}

The understanding of the electroweak symmetry breaking mechanisms represents a crucial step toward the completion of the Standard Model (SM) of particle physics, or toward its definitive dismissal to a low energy limit of some grander theory. This is the reason why the postulated Higgs boson\,\cite{Englert:1964et} has been sought after for the past 40 years at particle colliders, at ever increasing collison energy. The Tevatron collider produced the highest energy particle collisions up to 2009, and kept colliding protons and antiprotons until collecting approximately 10\,fb$^{-1}$ of data in 2011. 

Thanks to the distinctive proton-antiproton initial state, the CDF and D0 experiments provide unique physics to these days.
In particular, the Tevatron has measured with unprecedented precision some of the most important quantity of the Standard Model; the precise measurements of the top quark mass\,\cite{Tevatron:2014cka} and of the $W$ boson\,\cite{Aaltonen:2013iut} have been used over time to increasingly refine the indirect constraints on the mass of the SM Higgs boson\,\cite{Baak:2012kk}. The latest such fits are shown in Figure 1. Direct searches for the Higgs boson at the Tevatron have been performed for Higgs boson masses above the constraint set by the LEP experiments, leading to the first exclusion at higher mass range.
The fourth of July of 2012,  the ATLAS and CMS collaborations met the most significant milestone toward the search of  the Higgs boson: its observation!\,\cite{Aad:2012tfa,Chatrchyan:2012ufa}. The new particle mass was measured to be approximately 125\,GeV/$c^2$, consistent with SM predictions. The discovery was based on the decay of the new particle to known SM bosons: $\gamma \gamma, ZZ, W^+W^-$. At about the same time, the CDF and D0 collaborations teamed up to achieve the first evidence of the decay of the new particle\,\cite{Aaltonen:2012qt} into quarks. In 2014, the CMS and ATLAS collaboration both achieved strong direct evidence of Higgs boson coupling to leptons\,\cite{ATLAStautau,Chatrchyan:2014nva}.

 These proceedings refer to the talk given by the author on the result above and on the subsequent studies of the Higgs boson at proton-antiproton collisions. 

\begin{figure}[htb]
\centering
\includegraphics[height=3.7in]{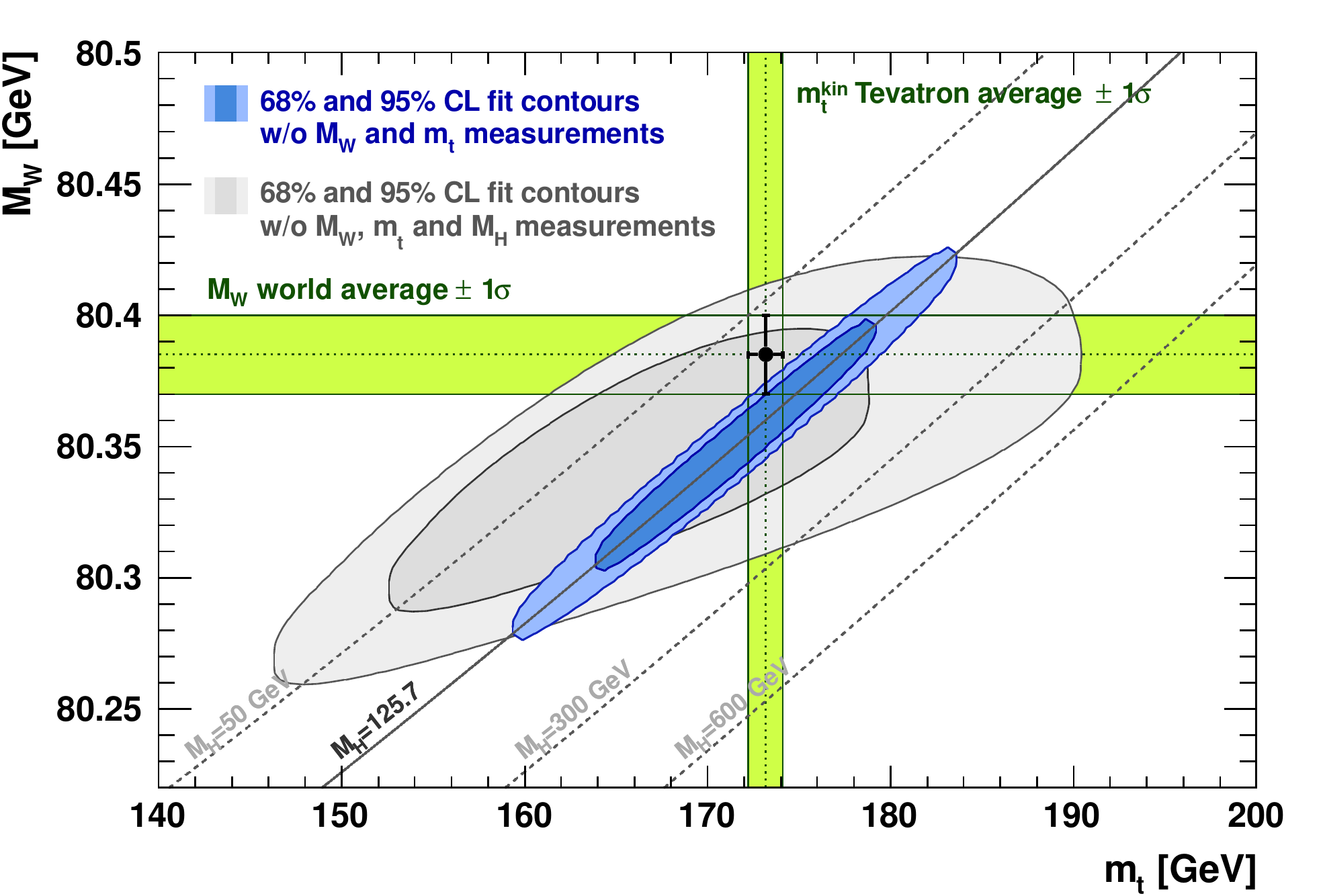}
\caption{Fits to global electroweak observables, projected onto the $M_{top}$ and $M_{W}$ axes. The green bands show the $68\%$ coverage bands around
the most precise $W$ and top quark mass determinations. The blue oval shows the Standard Model minimum. The new value for the measured top quark mass presented at this conference\,\cite{Petrillo} improves further the Standard Model internal consistency.}
\label{fig:figure1}
\end{figure}

\section{The Higgs Hunt at the Tevatron}

The strategy for the direct search for the Higgs boson at the Tevatron stems naturally from the composition of the different Higgs production and decay modes. The cross section for the dominant production modes at the Tevatron, i.e. direct Higgs production $gg\to H$, associated production of a Higgs boson with a gauge boson $q \bar q \to WH/ZH$,  and vector-boson fusion $q \bar q \to H q \bar q$ are shown in Table\,\ref{tab:table1}. In the same table are shown the branching ratio values for the decay modes studies at the Tevatron; all numbers assume a Higgs boson mass of 125\,GeV/$c^2$.

\begin{table}
\begin{center}
\begin{tabular}{lc|lc}  
Process                                &  Cross section (fb) &      Decay                              & Branching ratio ($\%$)\\ \hline
$gg\to H$                             &               950     &      $H \to b\bar b$                      &                  57.7  \\
$q \bar q \to WH$                &              130     &      $H \to W^+ W^-$                   &                21.6  \\
$q \bar q \to ZH$                 &                79      &      $H \to \tau^+ \tau^-$             &             6.4 \\
$q \bar q \to H q \bar q$     &                 67     &      $H \to \gamma \gamma$    &             0.2    \\
\end{tabular}  
\caption{Standard Model predictions for the main Higgs boson production modes, and Higgs decays, used in the Tevatron analyses. Most production cross sections have been computed at approximate next-to-next-leading order. Typical relative uncertainties on the branching rations are at the 5\% level.}
\label{tab:table1}
\end{center}
\end{table}

The CDF and D0 experiments are multi-purpose detector. While we refer to other documents to discuss in details the detectors features, we would like to stress the role of the silicon sensors, necessary to identify displaced secondary vertices originating from the in-flight decays of $B$ mesons, thus enabling the capability to search for the $H \to b \bar b$ decays.

\section{The Search Channels and Strategy}

The search for the Higgs boson at the Tevatron followed mainly a two-pronged strategy. At relatively low Higgs mass $m_H \le 135$\,GeV$c^2$ the most sensitive ways to probe for SM Higgs production is to focus on its associated production with vector bosons $W$ and $Z$. Data is selected from trigger path firing in the presence of a high-$P_T$ charged lepton, or large missing transverse energy ($MET$) - consistent with the leptonic decays of a vector boson - and requiring offline the presence of one or more $b$-jets as identified by the $b$-tagging algorithm, as expected by the SM decays of a low mass Higgs boson. At high mass $m_H \ge 135$\,GeV/$c^2$, the dominant production mechanism $gg\to H$ is exploited, together with the dominant decay of the Higgs to pairs of $W$ bosons in that mass region; the leptonic decays of both bosons allow for suppression of backgrounds, while limiting the amount of information that can be extracted in presence of a signal. To further increase sensitivity at masses at the edge of the "low" and "high" distinction, this analysis is expanded to include $WH \to WWW$ events in same-sign dilepton and trilepton final states.

In order to achieve sensitivity to SM production rates, an impressive amount of work went into refining the algorithms involved with every single piece of the analyses cascade: during Run\,II, the average sensitivity in each analysis channel improved by a factor 300\%, thus corresponding to a tenfold increase in integrated luminosity. In particular, acceptance is increased by better control over trigger thresholds through machine learning techniques, inclusion of loosely identified leptons, increasing efficiency in $b$-tagging identification algorithms\,\cite{Freeman:2012uf,Abazov:2013gaa}, complementing the identification of jets and missing transverse energy using information collected with the spectrometer in addition to the simple usage of calorimetric information\,\cite{Bentivegna:2012zj}, usage of machine-learning techniques to isolate the signal over the backgrounds exploiting kinematic and topological differences between the Higgs and other SM process, as well as their different correlations. The latter approach has been used either in a straightforward way, with one single discriminant against the sum of the backgrounds, or using a multi-staged approach when in presence of backgrounds kinematically very different among among them. 

In addition to the channels discussed above, the CDF and D0 collaborations performed analyses aimed at $H \to \tau \tau$ and $H \to \gamma \gamma$ decays. As the sensitivities reported are significantly less than the ones described above, these results are not covered in this talk. 
Both for high-mass and low-mass analysis, the same techniques have been used to measure the production rates of the smallest backgrounds in the same channels, thus providing additional validation of the tools applied. For low-mass analyses, an obvious choice - albeit a difficult one - was to measure the $WZ/ZZ$ production rates, in events where $Z \to b \bar b$ and the other boson decays leptonically. These events mimic closely the Higgs signal. The CDF and D0 collaborations measured in combination a cross section $\sigma(WZ+ZZ) = 3.0 \pm 0.6 (stat) \pm 0.7 (syst)$\,pb$^{-1}$\,\cite{ Aaltonen:2013kxa} in agreement with the SM expectation of $4.4 \pm 0.3$\,pb$^{-1}$. An interesting process with low cross section and same final state of the $WH \to \ell \nu b \bar b$ is the s-channel of the single top quark production $t \bar b \to W b \bar b \to \ell \nu b \bar b$\footnote{The charge conjugated processes are implicitly assumed throughout this paper. Unlike electroweak processes produced at the LHC, the cross sections for a process and its charge conjugated are identical at the Tevatron.}. The discovery of this process\,\cite{CDF:2014uma} is based on the usage of the tools applied in the $WH$ channel by D0 and CDF\,\cite{Abazov:2013qka,Aaltonen:2014qja}, and in the $VH \to MET b \bar b$ channel by CDF\,\cite{Aaltonen:2014xta}. At high mass, the $WW$ cross section has been measured with very large precision.

\section{Higgs Properties from Tevatron data}

The excess observed in Tevatron data has been analyzed in the hypothesis of pertaining to Higgs boson production, to infer on the nature of the Higgs boson itself. In particular, the excess has been compared to SM prediction to test for deviations with respect to its theoretical predictions; in particular, we report the measurements of the production rates, of the coupling of the Higgs boson to other SM elementary particles, and to the spin and parity structure of the new boson.

\subsection{Production Rates and Couplings}

\begin{figure}[htb]
\centering
\includegraphics[height=2.5in]{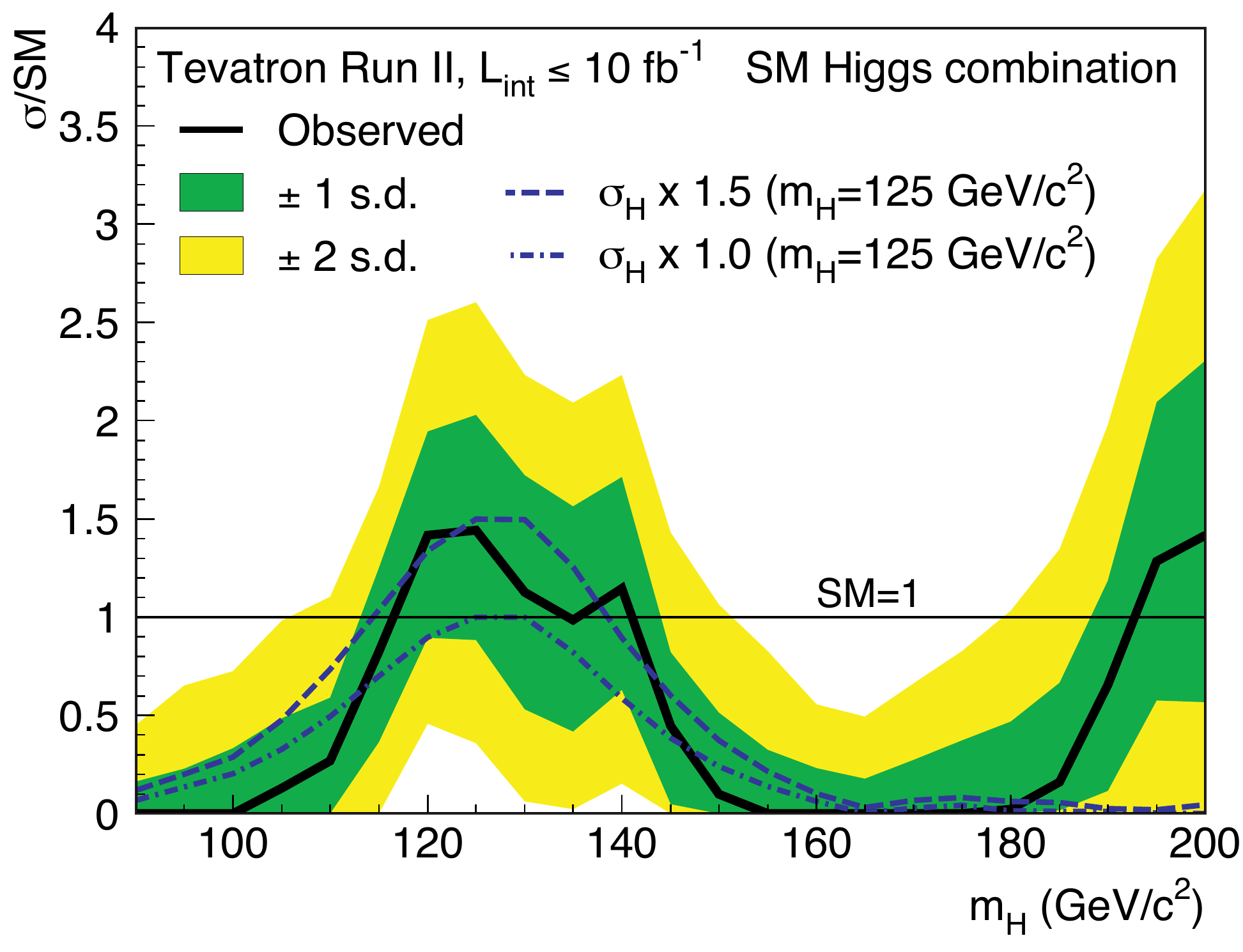}
\includegraphics[height=2.697in]{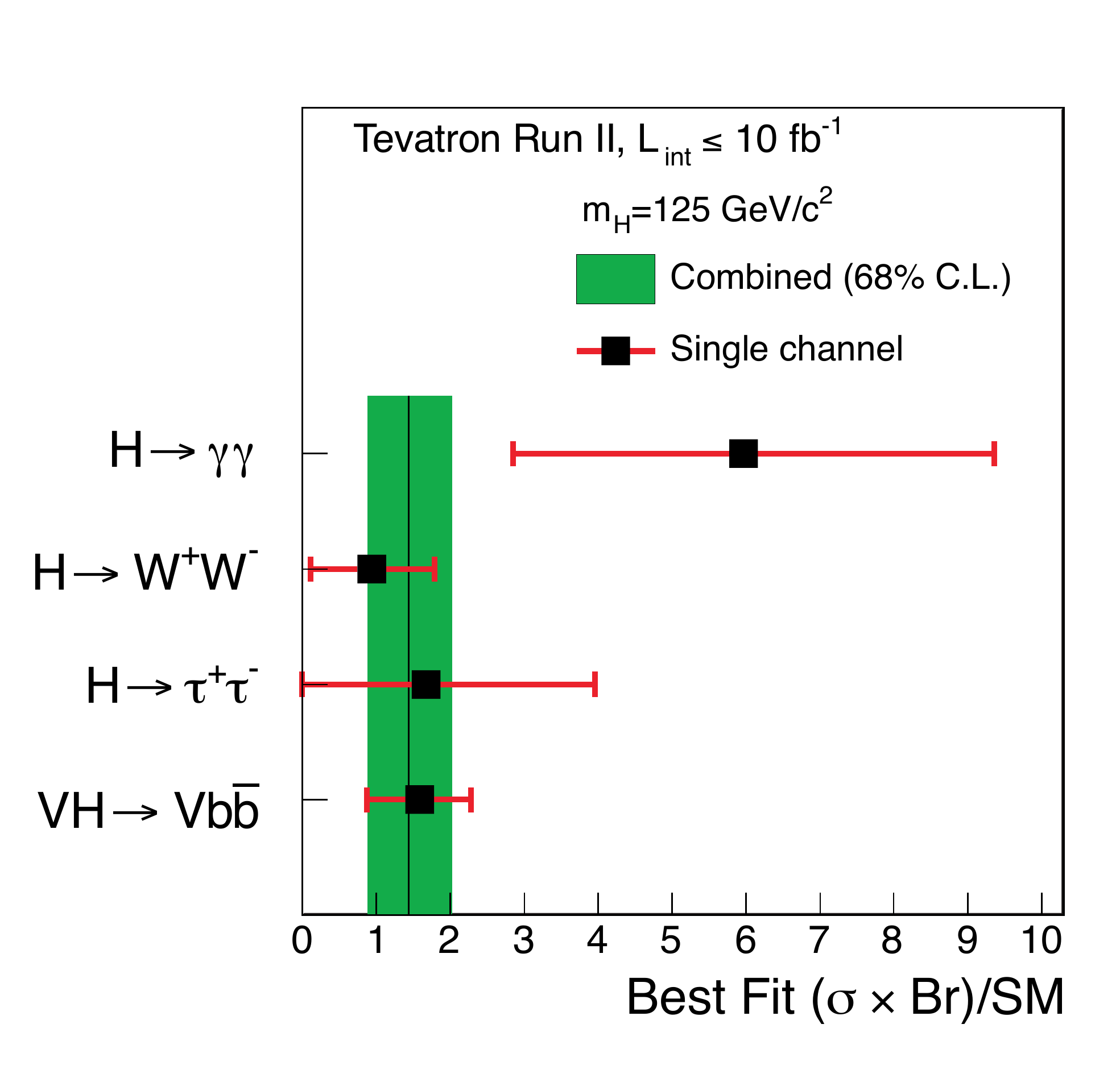}
\caption{The left plot shows the measured total signal strength $R = \frac{\sigma_{exp}} {\sigma_{SM}}$ 
assuming that the excess of events derive from a particle consistent with the Higgs boson. The right plot shows the signal strength in the analyzed decay modes, under the same assumption. The band on the right plot shows the measured $R$ with its uncertainties.}
\label{fig:figure2}
\end{figure}

Models with exotic couplings of the Higgs boson to other particles may enhance the production cross sections, the decay branching ratios, or the kinematic distributions of the signal. The overall production rate over SM prediction has been measured to be $R=1.44^{+0.59}_{-0.56}$ assuming a Higgs boson mass of 125\,GeV/$c^2$. The separate Higgs production cross section times branching ratio for $H \to \gamma \gamma$, $H \to WW$, $H \to \tau \tau$ and $H \to b \bar b$. The results for both fits are shown in Figure\,\ref{fig:figure2}. There is an excess in all analyzed channels. However, the main statistical significance comes from events compatible with the Higgs boson decaying to bottom quarks. Given that the LHC experiments do not see a large excess in the diphoton mass spectrum above SM Higgs production, while CDF and D0 do not expect to be sensitive to SM $H \to \gamma \gamma$, this excess should be interpreted as a statistical fluctuation of backgrounds. However, interpreting the global excess as signal will result in a shift in measured couplings. In particular, the $H \to \gamma \gamma$ final state is the only probe at the Tevatron of Higgs couplings that are sensitive to the relative sign of the Higgs coupling to bosons and fermions\footnote{The other one being single top and Higgs boson production for which only the LHC experiments will be sufficiently sensitive\,\cite{Biswas:2013xva}.}.

\subsection{Spin-parity}

Among the properties of the new boson that can be studied, are its spin and parity ($J^P$) structure. The ATLAS and CMS collaborations have tested non-SM spin-parity hypotheses using the $H \to ZZ, H \to WW $ and $H \to \gamma \gamma$ decay modes. The Tevatron data allows the unique exploration of the spin-parity structure in the $H \to b \bar b$ decays. Different $J^P$ assumptions imply cross sections that rise with the velocity $\beta$ of the Higgs boson, with the SM assignment $J^P=0^+$ leading to a dependence of the cross section linearly on $\beta$, the pseudo scalar $J^P=0^-$ hypothesis leading to a dependence on $\beta^3$ and a graviton-like assignment $J^P=2^+$ leading to a dependence to the fifth power $\beta^5$\,\cite{Ellis:2012xd}.
The two non-SM hypotheses thus predict large variations in the signal kinematics, that can be probed using existing data. The single most striking difference with respect to SM predictions would be present in the invariant mass of all visibile object, an observable that is a proxy for $\sqrt{s}$. The D0 re-analysis of the same data used for the search for $H \to b \bar b$ exclude the interpretation of the Higgs excess in the non-SM spin-parity hypothesis assuming SM cross sections, at approximately three standard deviations, for both the pseudo scalar and the graviton-like hypotheses\,\cite{Abazov:2014doa}. The D0 collaboration also sets upper limits on the fraction of non-SM signal in the data under the assumption of two nearly degenerate bosons with different spin and parity values.

After this conference, the CDF Collaboration released a similar re-analysis of $H \to b \bar b$ channels to provide independent exclusions of non-SM spin-parity hypothesis using fermionic Higgs decays\,\cite{CDFJP} at comparable sensitivities. Finally, CDF reanalyzed its data to search for invisible decays of the Higgs boson; these results are not competitive with LHC ones and are thus not discussed in detail here.

\begin{figure}[htb]
\centering
\includegraphics[height=2.64in]{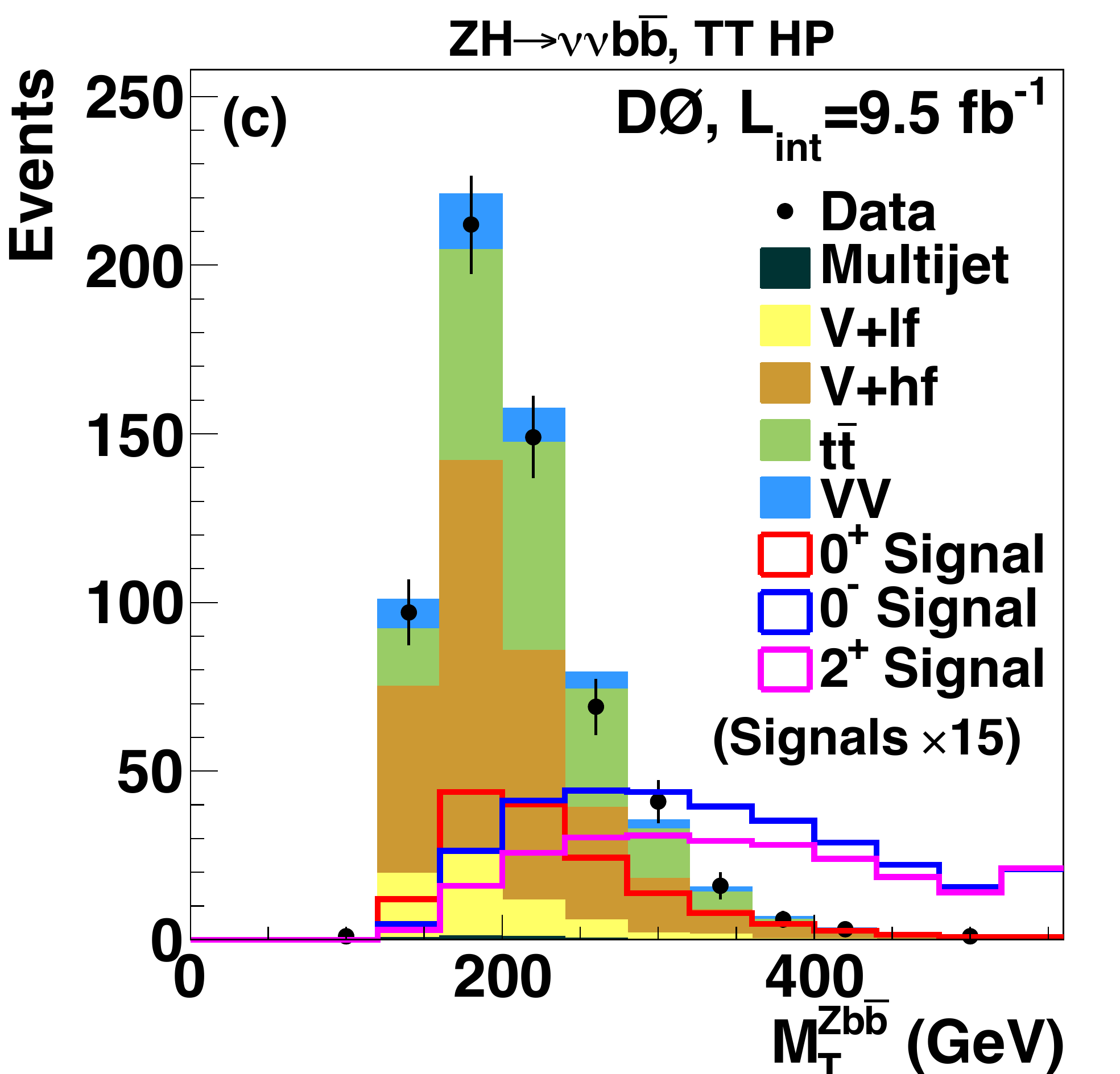}
\includegraphics[height=2.5in]{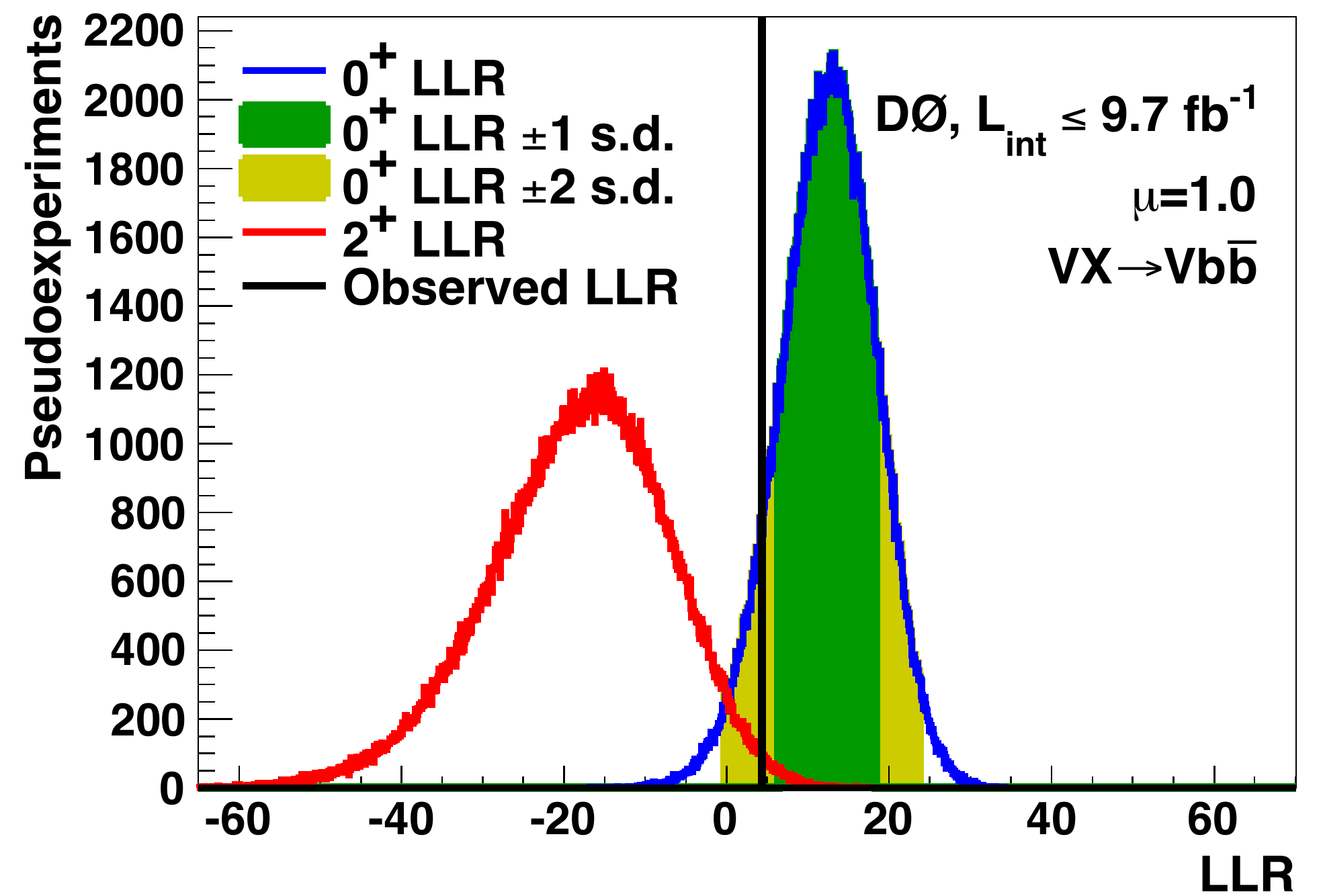}
\caption{The left plot shows the distribution of the invariant mass of all visible objects for the analysis targeting $ZH \to MET b \bar b$ events, and how this distribution would be distorted by exotic assignments of the spin and parity of the new boson. The $J^P = 0^?$ and $J^P = 2^+$ samples are normalized to the product of the SM cross section and branching fraction, multiplied by an additional x15 factor to enhance visibility.
The right plot shows the test-statistics in the assumption that the new boson is produced according with SM cross sections. The vertical solid line represents the observed LLR value assuming $\mu = 1.0$. }
\label{fig:figure3}
\end{figure}

\section{Additional Remarks}

The results highlighted in this talk, show how the distinctive initial state configuration of the Tevatron collider allowed for exploration of the electroweak symmetry breaking mechanism in channels complementary to the ones explore at the LHC. 
It should be noted that it will be challenging to further improve LHC sensitivity to $H \to b \bar b$ the coming run, unless new theoretical and technological advances are put forward. An alternative way to probe Higgs coupling to bottom quarks would be studying the production of the Higgs boson together with bottom quarks $b \bar b H$. However, the analysis of this channel would likely be even more challenging than studying Higgs coupling to bottom quarks in the Higgs decays. 

The Tevatron experiment leveraged on the better S/B ratio in the signature with $b-$jets and leptons to produce sensitive searches for new physics in the electroweak sector.  One such example is CDF the search for an exotic $W^{\prime}$ boson in the $t \bar b$ final state\,\cite{WprimeCDF}.
 The selection of the leptonic decays chain of the top quark ($W^{\prime} \to t \bar b \to W b \bar b \to \ell \nu b \bar b$) gives the same final state as in $WH \to \ell \nu b \bar b$ although with different kinematics. The analysis presented at this conference adds for the first time the events where the charged lepton has not been explicitly identified. An analysis that includes identified leptons has been approved after this conference and combined with the one shown here to provide the most stringent constraints on this model\,\cite{WprimeCDF_new}.
 While a $W^{\prime}$ with the same coupling as a SM $W$ boson has already been excluded at intermediate masses (up to 800\,GeV/$c^2$) with the Tevatron data, and beyond ($\ge 800$\,GeV/$c^2$) by LHC data, the new result allow to set the tightest constraints on the coupling of this hypothetical new particle for $W^{\prime}$ masses below 650\,GeV/$c^2$.

A similar situation occurs for the search for a new neutral boson $Z^{\prime}$ decaying to $t \bar t$. Since this new particle does not couple directly to gluons, its cross section increases only mildly at the LHC energies, while the irreducible $t \bar t$ background increases at  a much larger rate. A CDF analysis of the full Tevatron dataset searches for a resonant $t \bar t \to \ell \nu + (b)jets$ setting the strongest constraints on $Z^{\prime}$ couplings for masses below 750\,GeV/$c^2$\,\cite{Aaltonen:2012af}.

The points highlighted above should be kept in mind in the scenario where new physics in the electroweak sector could manifest itself at relatively low energies in the next years. It should be thus clear how relevant is the ongoing effort to preserve CDF and D0 data and their access to future users.

\section{Conclusions}

In this talk we discussed the role of the CDF and D0 experiments in understanding the mechanism of electroweak symmetry breaking and mass generation for fermions. An excess consistent with the production of a Higgs boson decaying mostly to bottom quarks is present in CDF and D0 data, with a combined significance corresponding to approximately three standard deviations. The data is used to set constraints on the coupling of the new boson to SM particles, and to test the spin-parity structure of the Higgs boson. The data for both coupling and spin-parity analyses favor the SM Higgs interpretation within the experimental uncertainties.

\Acknowledgements
The author wishes to thank the conference organizers for the excellent organization, and for providing a stimulating ground for discussions. Special thanks go the the D0 and CDF collaborators who have dedicated their scientific careers to the understanding of the coupling of the Higgs boson to quarks.

\end{document}

%% file: econfmacros.tex
\def\beq{\begin{equation}}
\def\eeq#1{\label{#1}\end{equation}}
\def\eeqn{\end{equation}}

\def\beqa{\begin{eqnarray}}
\def\eeqa#1{\label{#1}\end{eqnarray}}
\def\eeqan{\end{eqnarray}}

\let\bar=\overbar

\def\Dslash{\not{\hbox{\kern-4pt $D$}}}
\def\dslash{\not{\hbox{\kern-2pt $\del$}}}

\def\msb{{\bar{\ssstyle M \kern -1pt S}}}




%% file: Margaroli_LHCP.bbl
\newpage

\begin{thebibliography}{99}


\bibitem{Englert:1964et} 
  F.~Englert and R.~Brout,
  Phys.\ Rev.\ Lett.\  {\bf 13}, 321 (1964).
  
  P.~W.~Higgs,
  Phys.\ Rev.\ Lett.\  {\bf 13}, 508 (1964).
  
  G.~S.~Guralnik, C.~R.~Hagen and T.~W.~B.~Kibble,
  Phys.\ Rev.\ Lett.\  {\bf 13}, 585 (1964).


\bibitem{Tevatron:2014cka} 
  Tevatron Electroweak Working Group [CDF and D0 Collaborations],
  arXiv:1407.2682


\bibitem{Aaltonen:2013iut} 
  T.~A.~Aaltonen {\it et al.}  [CDF and D0 Collaborations],
  Phys.\ Rev.\ D {\bf 88}, no. 5, 052018 (2013)
  arXiv:1307.7627


\bibitem{Baak:2012kk} 
  M.~Baak, M.~Goebel, J.~Haller, A.~Hoecker, D.~Kennedy, R.~Kogler, K.~Moenig and M.~Schott {\it et al.},
  Eur.\ Phys.\ J.\ C {\bf 72}, 2205 (2012)
  arXiv:1209.2716
  
  
  \bibitem{Aad:2012tfa} 
  G.~Aad {\it et al.}  [ATLAS Collaboration],
  Phys.\ Lett.\ B {\bf 716}, 1 (2012)
  arXiv:1207.7214
  
  
\bibitem{Chatrchyan:2012ufa} 
  S.~Chatrchyan {\it et al.}  [CMS Collaboration],
  Phys.\ Lett.\ B {\bf 716}, 30 (2012)
  arXiv:1207.7235

\bibitem{Aaltonen:2012qt} 
  T.~Aaltonen {\it et al.}  [CDF and D0 Collaborations],
  Phys.\ Rev.\ Lett.\  {\bf 109}, 071804 (2012)
  arXiv:1207.6436

\bibitem{ATLAStautau}
G.~Aad {\it et al.}  [ATLAS Collaboration],
ATLAS-CONF-2013-108


\bibitem{Chatrchyan:2014nva} 
  S.~Chatrchyan {\it et al.}  [CMS Collaboration],
  JHEP {\bf 1405}, 104 (2014)
  arXiv:1401.5041


  
  \bibitem{Petrillo}
  G.~Petrillo, these proceedings
  
  
  
\bibitem{Freeman:2012uf} 
  J.~Freeman, T.~Junk, M.~Kirby, Y.~Oksuzian, T.~J.~Phillips, F.~D.~Snider, M.~Trovato and J.~Vizan {\it et al.},
  Nucl.\ Instrum.\ Meth.\ A {\bf 697}, 64 (2013)
  arXiv:1205.1812
  
\bibitem{Abazov:2013gaa} 
  V.~M.~Abazov {\it et al.}  [D0 Collaboration],
  arXiv:1312.7623
  
\bibitem{Bentivegna:2012zj} 
  M.~Bentivegna, Q.~Liu, F.~Margaroli and K.~Potamianos,
  arXiv:1205.4470
  
  
  
  
\bibitem{Aaltonen:2013kxa} 
  T.~Aaltonen {\it et al.}  [CDF and D0 Collaborations],
  Phys.\ Rev.\ D {\bf 88}, no. 5, 052014 (2013)
  arXiv:1303.6346
  
  
\bibitem{CDF:2014uma} 
  T.~A.~Aaltonen {\it et al.}  [CDF and D0 Collaborations],
  Phys.\ Rev.\ Lett.\  {\bf 112}, 231803 (2014)
  arXiv:1402.5126
  
\bibitem{Abazov:2013qka} 
  V.~M.~Abazov {\it et al.}  [D0 Collaboration],
  Phys.\ Lett.\ B {\bf 726}, 656 (2013)
  arXiv:1307.0731
  
\bibitem{Aaltonen:2014qja} 
  T.~A.~Aaltonen {\it et al.}  [CDF Collaboration],
  Phys.\ Rev.\ Lett.\  {\bf 112}, 231804 (2014)
  arXiv:1402.0484
  
\bibitem{Aaltonen:2014xta} 
  T.~A.~Aaltonen {\it et al.}  [CDF Collaboration],
  Phys.\ Rev.\ Lett.\  {\bf 112}, 231805 (2014)
  arXiv:1402.3756
  
\bibitem{Biswas:2013xva} 

  S.~Biswas, E.~Gabrielli and B.~Mele,
  JHEP {\bf 1301}, 088 (2013)
  arXiv:1211.0499

  M.~Farina, C.~Grojean, F.~Maltoni, E.~Salvioni and A.~Thamm,
  JHEP {\bf 1305}, 022 (2013)
   arXiv:1211.3736

  S.~Biswas, E.~Gabrielli, F.~Margaroli and B.~Mele,
  JHEP {\bf 07}, 073 (2013)
  arXiv:1304.1822
  
  
  
\bibitem{Ellis:2012xd} 
  J.~Ellis, D.~S.~Hwang, V.~Sanz and T.~You,
  JHEP {\bf 1211}, 134 (2012)
  arXiv:1208.6002
  
\bibitem{Abazov:2014doa} 
  V.~M.~Abazov {\it et al.}  [D0 Collaboration],
  arXiv:1407.6369
  
\bibitem{CDFJP} 
T.~Aaltonen {\it et al.} [CDF Collaboration] 
CDF Conference Note 11103
  

  
  \bibitem{WprimeCDF}
   T.~Aaltonen {\it et al.}  [CDF Collaboration],
   CDF Conference Note 11079
   
  \bibitem{WprimeCDF_new}
  T.~Aaltonen {\it et al.}  [CDF Collaboration],
  CDF Conference Note 11110 

\bibitem{Aaltonen:2012af} 
  T.~Aaltonen {\it et al.}  [CDF Collaboration],
  Phys.\ Rev.\ Lett.\  {\bf 110}, no. 12, 121802 (2013)
  arXiv:1211.5363
  
\end{thebibliography}
